# Complementary bodies in sphere packing

Philip W. Kuchel

*Faculty of Science, University of Sydney, New South Wales 2006, Australia*

*e-mail: philip.kuchel@sydney.edu.au*

Symbolic and graphical tools, such as *Mathematica*, enable precise visualization and analysis of void spaces in sphere packings. In the cubic close packing (CCP, or face-centred cubic packing; FCC) arrangement these voids can be partitioned into repeating geometric units we term spherically truncated polyhedra - bodies analogous to plane-truncated polyhedra but bounded by both planar and spherical surfaces. These structures are relevant in geometric studies and applications such as modelling diffusion in porous media and biological tissues. This work examines the properties of these complementary bodies, deriving their surface area-to-volume ratios, which are significant in physical contexts; and we establish a result concerning the packing density of truncated tetrahedra and octahedra, demonstrating how they tile the interstitial space surrounding packed spheres. These findings contribute to a deeper understanding of classical packing problems and their geometrical complements.

## 1. Introduction

Traditional studies of sphere packing primarily focus on the arrangement and efficiency of sphere placement. However, our work here shifts perspective by investigating the geometry of the interstitial voids - the spaces between spheres - and reconstructing the packing structure from these complementary bodies. In particular, we examine the void spaces within the cubic close packing (CCP) arrangement, demonstrating how they can be partitioned into fundamental repeating units, termed spherically truncated polyhedra. These units fully tile the voids, with the spherical cavities emerging naturally from their assembly.

To illustrate this approach, we begin with a physical model: a pyramid of steel spheres arranged in CCP (Fig. 1). The small, cuspate chambers between adjacent spheres define the unit voids of interest. Unlike traditional crystallographic descriptions that emphasize lattice points [1], this method directly examines the geometry of void spaces. Previous work by Graton and Fraser (1935) [2] laid a foundation by using hand-drawn diagrams to study voids in geological sphere packings. However, their empirical treatment lacks the formal precision required for modern applications in porous material modelling.

These void structures hold significance in multiple fields, including estimating dielectric and magnetic field gradients in heterogeneous materials [3–5], modelling drug binding in viscous media [6], and simulating diffusion in porous systems [4, 8–10]. Our interest in these void spaces arises particularly from nuclear magnetic resonance (NMR) experiments, where restricted diffusion in packed sphere arrays results in periodic modulations of signal intensity as a function of magnetic field gradient strength [11–13].

A crucial parameter in such systems is the packing density of spheres. Kepler famously conjectured in 1611 that the maximal density for uniform sphere packings is $\frac{\pi}{\sqrt{18}}$, a claim formally proven in 2014 [7]. This article takes CCP as a canonical model to explore void space geometry within ordered materials. The discussion proceeds in three parts:

1. *Graphical analysis* using *Mathematica*'s advanced visualization tools to construct representations of spherically truncated polyhedra.
2. *Derivation of geometric properties*, including surface area and volume expressions.
3. *Tiling results*, demonstrating how two specific truncated polyhedral types combine to fill the void space with the same density as the original CCP sphere arrangement.

These insights contribute to the broader understanding of packing structures and their geometric complements, offering a refined perspective on void space organization within ordered sphere arrays.

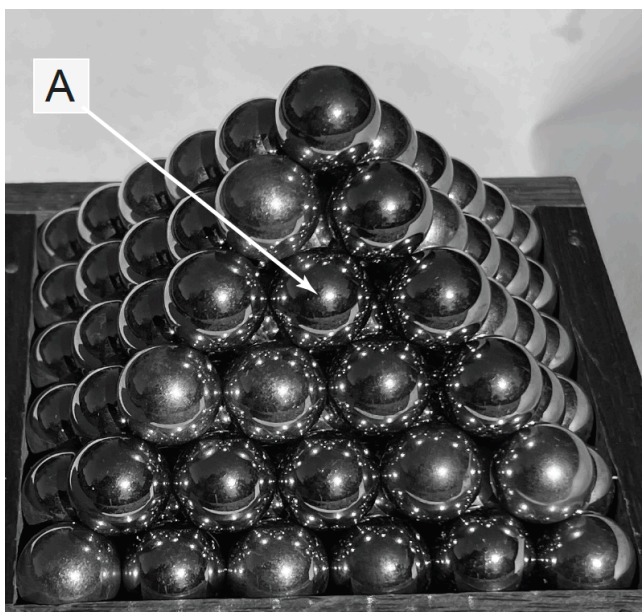

**Fig. 1.** Oblique view of 0.5-inch (12.68 mm) diameter steel spheres in CCP. The shape of the pyramid is essentially the top half of a regular octahedron; each sloping face is an equilateral triangle formed by 1 + 2 + 3 + 4 + 5 + 6 spheres. **A** is directed to a sphere that is surrounded on the face by six others, with six cuspate depressions (term used by [2]) of two different forms surrounding it; the significance of these is described below.

## 2. Shapes of Unit Voids in CCP

### 2.1. Overview

Using *Mathematica*'s advanced graphics capabilities, arrays of tangential spheres were rendered and made selectively transparent to reveal the *unit voids* - the distinct spaces between spheres. By removing the spheres, these voids become standalone geometric objects suitable for analysis (URL for *Mathematica* Community link to Notebook).

Following the approach of Graton and Fraser [2], the total void can be partitioned using planes through sphere centres. They showed that only two distinct types of unit voids are needed to tile the entire void space. Figure 1 illustrates this insight, showing depressions of two forms around a central sphere - shallower depressions capped by a fourth tangential sphere (tetrads), and deeper ones forming part of a six-sphere group (hexamers).

To investigate further, we constructed a symmetric virtual CCP array of 19 spheres (Fig. 2), centred at the origin. This compact configuration preserves the features of larger CCP stacks and allows detailed visualization of interstitial geometries. Within this structure, hexamers of overlapping spheres form octahedral frameworks, whose edges mark points of contact.

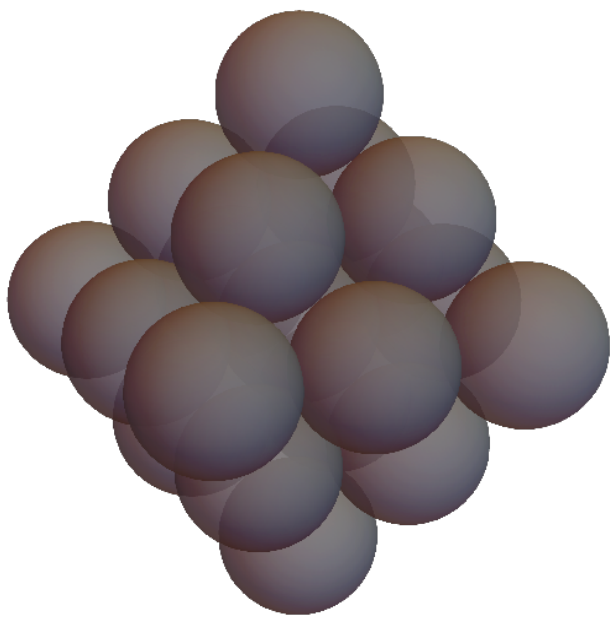

**Fig. 2.** *Mathematica*-rendered array of spheres (radius R = 1 unit) based on Fig. 1 that was translated so that one sphere was centred at the origin, and the other centres listed from left to right and front to back, became: {2, -2, 0}, {2, 0, 0}, {2, 2, 0}, {0, -2, 0}, {0, 0, 0}, {0, 2, 0}, {-2, -2, 0}, {-2, 0, 0}, {-2, 2, 0}, {1, -1, $\sqrt{2}$}, {1, 1, $\sqrt{2}$}, {-1, -1, $\sqrt{2}$}, {-1, 1, -$\sqrt{2}$}, {0, 0, 2$\sqrt{2}$}, {1, -1, -$\sqrt{2}$}, {1, 1, -$\sqrt{2}$}, {-1, -1, -$\sqrt{2}$}, {-1, 1, -$\sqrt{2}$}, {0, 0, -2$\sqrt{2}$}.

## 2.2. Spherically truncated octahedron (STO)

A hexamer is readily visualised by starting with the top sphere in Fig. 2, its four neighbours below it, and then the central sphere below them. This forms the framework of the configuration shown in Fig. 3.

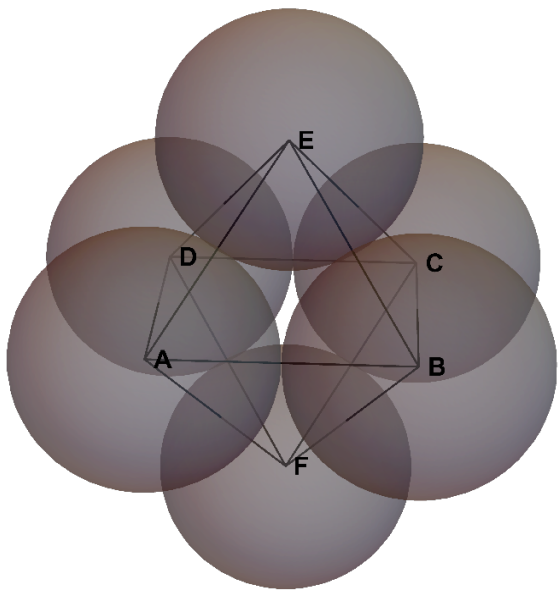

**Fig. 3.** Hexamer of identical spheres, each tangential to four others. In the centre of the projected image is what appears to be a curve-sided triangle in front of one of the same shape behind, rotated by 180º, creating a crossed triangle appearance that is apparent on close inspection in three instances around **A** in Fig. 1. The regular octahedron formed from vertices at the centres of each sphere is labelled **A**,…, **F**, and the midpoints of the edges of the octahedron are the points of contact between neighbouring spheres.

The six-sphere hexamers define regular octahedra. When these are truncated at their corners by spherical caps centred on each vertex, the resulting void shape - bounded by 6 spherical squares and 8 curvilinear triangular faces - is what we call the spherically truncated octahedron (STO). This shape is analogous to a cuboctahedron but with curved faces.

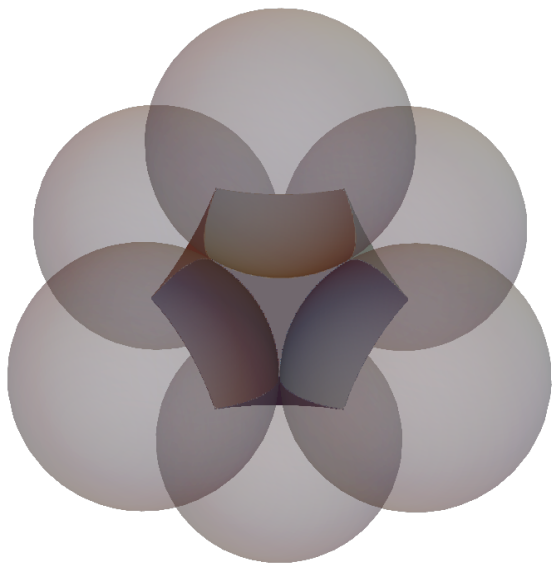

**Fig. 4.** Spherically truncated octahedron (STO; also called a sphero-planar cuboctahedron; dark grey) completely filling the void between its six generating CCP spheres.

The STO was rendered as a GraphicsComplex in *Mathematica* and can be repositioned in space using various commands. Docking six STOs around a central point (Fig. 5) reveals the presence of additional voids not filled by STOs alone.

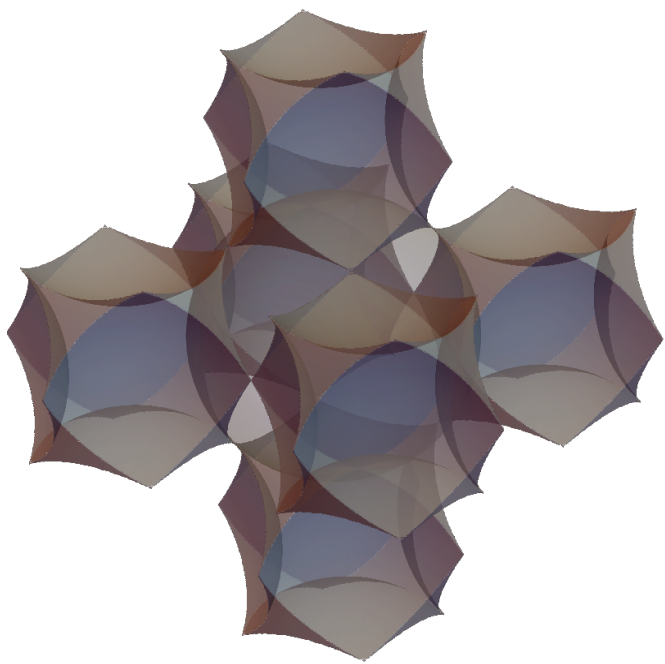

**Fig. 5.** Six STOs (dark grey) that fit in the unit voids around a central sphere (light grey) such as in Fig. 2.

### 2.3. Spherically truncated tetrahedron (STT)

The remaining voids correspond to the tetrads formed by four mutually tangential spheres (Fig. 6).

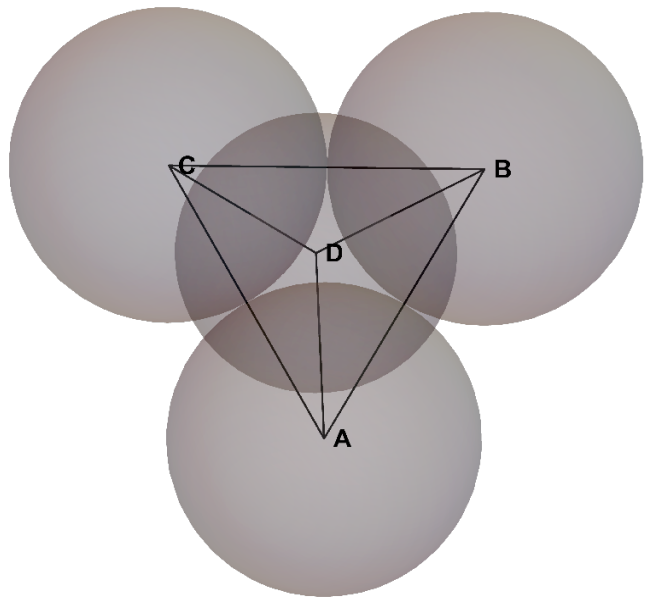

**Fig. 6.** Four mutually tangential spheres from the face of Fig. 2 whose centres form the vertices of a tetrahedron.

Their centres define a tetrahedron. Truncating this volume with spherical caps from each sphere yields a new shape: the spherically truncated tetrahedron (STT), which was also visualized using *Mathematica* (Fig. 7).

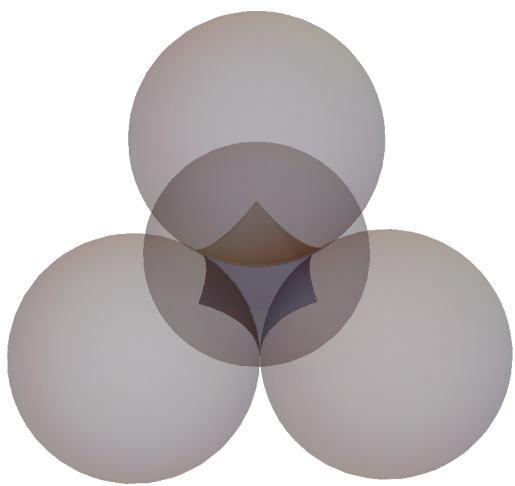

**Fig. 7.** Spherically truncated tetrahedron (STT; also called a sphero-planar octahedron) that fills the unit void between four mutually tangential spheres.

## 2.4. Complete tiling

By docking eight STTs to the six STOs around a central sphere - rotating opposing STTs by 90° to achieve a snug fit - the entire void surrounding a central sphere is filled (Fig. 8). This configuration shows that CCP void space can be perfectly tiled by a combination of STOs and STTs.

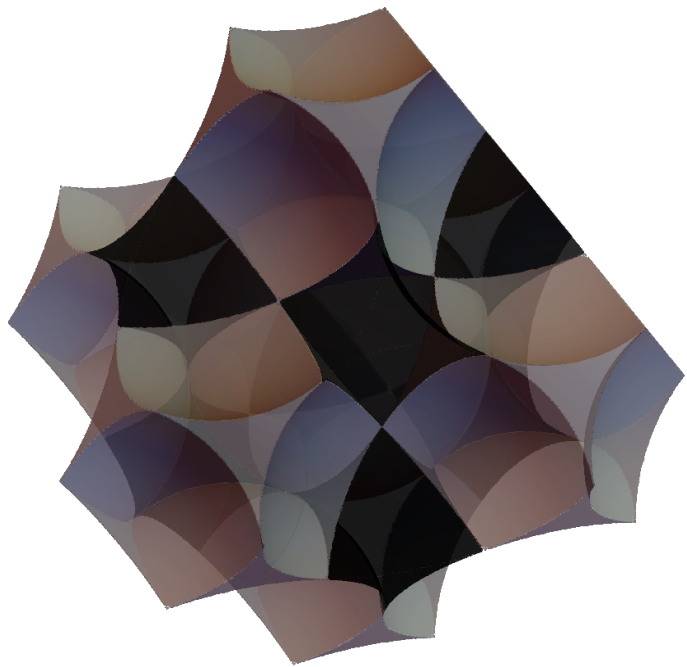

**Fig. 8.** Six STOs and eight STTs totally enclosing a central sphere that is part of their mathematical generating function.

## 3. Mathematical Properties

We now derive expressions for the surface area and volume corresponding to the geometric constructions introduced above. In the case of the sphero-planar polyhedra, the surface consists of two distinct types of region: (1) *convex* faces, which are portions of the truncating spheres; and (2) *planar* triangular faces, which lie in Euclidean planes defined by adjacent sphere centres and are bounded by concave circular arcs. To compute the area of the convex spherical faces, we use the theory of spherical triangles [16–18], which allows us to determine the area of a triangle inscribed on the surface of a sphere. Figure 9 illustrates such a spherical triangle, labelled ABC, with side lengths and angles subtended at the sphere's centre, O, formed by radii extending from O to each of the triangle's vertices.

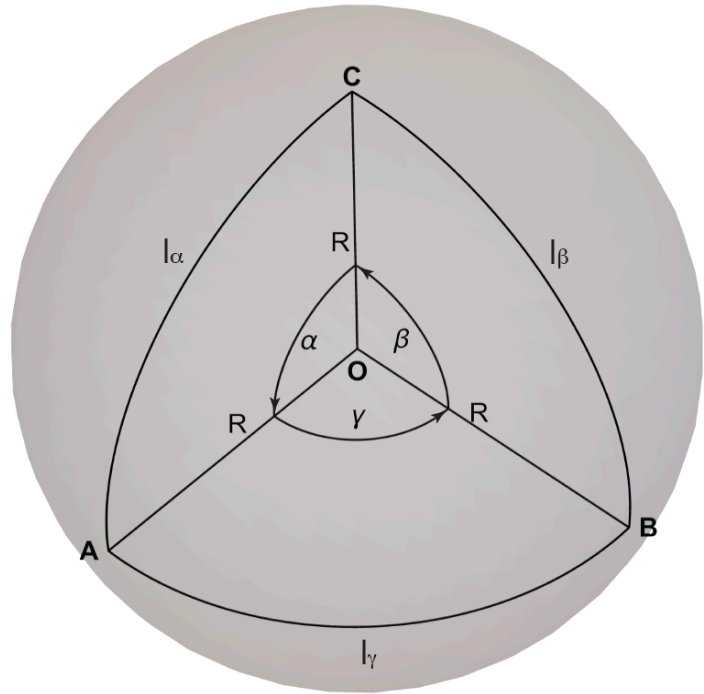

**Fig. 9.** Spherical triangle drawn on the surface of a sphere of radius R, with side (arc) lengths $l_\alpha$, $l_\beta$, and $l_\gamma$ that subtend the respective angles $\angle AOC = \alpha$, $\angle BOC = \beta$, and $\angle AOB = \gamma$ at the centre of the sphere.

## 3.1. STT surface area ($A_{STT}$)

In Figure 9, uppercase Roman letters denote the internal angles of the spherical triangle, while lowercase Greek letters represent the corresponding angles subtended at the centre of the sphere. The area of the spherical triangle is determined using its *spherical excess*, E [16–18], which quantifies how much the sum of the triangle's angles exceeds that of a planar triangle.

The spherical excess is defined as:

$$E = (\angle A + \angle B + \angle C - \pi). \tag{1}$$

Using the cosine rule for spherical triangles [19] with C denoting the angle at the vertex C in the spherical triangle:

$$\cos \gamma = \cos \alpha \cos \beta + \sin \alpha \sin \beta \cos C. \tag{2}$$

Setting $\alpha = \beta = \gamma = \pi/3$ in equation (2) gives $\cos C = 1/3$. Then the spherical excess is:

$$E = 3\,C - \pi \ . \tag{3}$$

Using the triple-angle addition formula for cosine [*e.g.*, 20] gives cos E = 23/27 so

$$E = \arccos(23/27) = 0.551286 \ . \tag{4}$$

The area, A, of a spherical triangle is simply [16-18]:

$$A = R^2\,E \ . \tag{5}$$

With a side-length of 2 R (R = 1) the area of a face of the original tetrahedron is $\sqrt{3}$ (while the area of the three sectors on the face is $\frac{\pi}{2}$, so the area of the four planar concave triangular faces of the STT, P, is:

$$P = 4\,\sqrt{3} - 2\,\pi \ . \tag{6}$$

With four planar concave triangles and four spherical sectors (spherical triangles) on each STT the contribution, Q, to its whole area is 4 E. Thus, the total surface area of the STT, $A_{STT}$, is:

$$A_{STT} = 4\,E + P\,, \tag{7}$$

$$A_{STT} = 4\arccos\left(\tfrac{23}{27}\right) + \left(4\sqrt{3} - 2\pi\right). \tag{8}$$

This evaluates to 2.85016, and relative to the area of one of the generating spheres it is 0.226809.

### 3.2. STT volume ($V_{STT}$)

From equation (4) the volume of each spherical sector, J, is given by the ratio of a spherical triangle divided by the surface area of the sphere, times the volume of the sphere:

$$J = \frac{\arccos\left(\frac{23}{27}\right)}{4\pi} \frac{4\pi}{3} = \frac{1}{3} \arccos\left(\frac{23}{27}\right) . \tag{9}$$

The volume of a regular tetrahedron, K, of side-length a is:

$$K = \frac{a^3}{6\sqrt{2}} , \tag{10}$$

and with a side length of 2 R (R = 1):

$$K = \frac{2\sqrt{2}}{3} . \tag{11}$$

Finally, the volume of an STT, $V_{STT}$, is 4 J (equation 8) subtracted from K (equation 10):

$$V_{STT} = \frac{2}{3}\left(\sqrt{2} - 2\arccos\left(\frac{23}{27}\right)\right) , \tag{12}$$

which evaluates to 0.207762. Thus, the volume of an STT relative to one of its generating spheres is:

$$\frac{V_{STT}}{\left(\frac{4\pi}{3}\right)} = 0.0495994 , \tag{13}$$

namely, ~5% of the volume of a generating sphere.

### 3.3. STT sphere-section packing density, $\phi_{STT}$

The four spherical sectors, 4 J, that are removed from the tetrahedron (equation 8) in generating an STT, have a total volume relative to the whole tetrahedron that is given by,

$$\phi_{STT} = \sqrt{2} \arccos\left(\frac{23}{27}\right) . \tag{14}$$

This evaluates to,

$$\phi_{STT} = 0.779636 . \tag{15}$$

The significance of the above values is considered more below.

Turn now to the STO.

### 3.4. STO surface area ($A_{STO}$)

The STO analysis follows that of the STT, but the spherical triangles of the STT are replaced by spherical squares. A spherical square is composed of two spherical triangles, formed by diagonal bisection; and applying the cosine law for spherical triangles (equation 2, [19]) with $\alpha = \beta = \pi/3$ and $\gamma = \pi/2$ gives the area of each half square as arccos(7/9); hence the total area of each spherical square is given by the spherical excess (with R = 1):

$$E = 2 \arccos\left(\frac{7}{9}\right) = \arccos(17/81) \quad . \tag{16}$$

The area of a planar concave triangle, $A_{CT}$, on each of the eight faces of the octahedron (Fig. 5) is:

$$A_{CT} = \left(\sqrt{3} - \frac{\pi}{2}\right) . \tag{17}$$

Therefore, from equations (16) and (17), the total surface area of an STO (recall R = 1) is:

$$A_{STO} = 6\,E + 8\,A_{CT}$$

$$= 6 \arccos\left(\frac{17}{81}\right) + 8\left(\sqrt{3} - \frac{\pi}{2}\right) . \tag{18}$$

Numerically $A_{STO}$ = 9.44612, and relative to one of the generating spheres (area = $4\pi$) this is 0.751698.

### 3.5. STO volume ($V_{STO}$)

The volume of each spherical sector, P, is given by the ratio of the area a spherical square divided by the surface area of the sphere, times the volume of the sphere:

$$P = \frac{\arccos(\frac{17}{81})}{4\pi} \frac{4}{3}\pi = \frac{1}{3}\arccos\left(\frac{17}{81}\right) . \tag{19}$$

The volume of a regular octahedron of side-length a is:

$$Q = \frac{\sqrt{2}\, a^3}{3}, \tag{20}$$

and with two sphere radii (R = 1) on each edge this gives a = 2 so:

$$Q = \frac{\sqrt{2}\, 8}{3}. \tag{21}$$

Therefore, since there are six vertices that are truncated from the parent octahedron:

$$V_{STO} = Q - 6\,P = \frac{\sqrt{2}\, 8}{3} - 2\arccos\!\left(\frac{17}{81}\right). \tag{22}$$

This evaluates to 1.05254 which, relative to the volume of one of the generating spheres (with $R = 1$; $4\pi/3 = 4.18879$), is 0.251276.

### 3.6. STO sphere-section packing density, $\phi_{STO}$

The total volume of the sphere sectors, 6 P, that are removed from the octahedron (equation 19) in generating an STT has a volume relative to the whole tetrahedron (equation 22) that is:

$$\phi_{STO} = \frac{2\arccos\!\left(\frac{17}{81}\right)}{\frac{\sqrt{2}\,8}{3}} = \frac{3}{4\sqrt{2}}\arccos\!\left(\frac{17}{81}\right) = 0.720903. \tag{23}$$

### 3.7. STT and STO volume ratio

The volume of an STT relative to an STO is 0.197391 or ~20%.

## 4. Conjecture of STT and STO Packing

As noted in the Introduction, the maximum packing density for an unbounded array of uniform spheres is $\pi/\sqrt{18} = 0.74048$ [7]. But it can be seen from equation (18) that $\phi_{STT}$ is greater than this, while from equation (28), $\phi_{STO}$ is less.

Numerical experiments with a *Mathematica* script (not shown here; see the abovementioned URL link) revealed that the weighted average given by (2 ϕ$_{STO}$ + 1 ϕ$_{STT}$) /3 was 0.0984697… which equates to π/√18 to at least 14 decimal places. However, this is merely a numerical result and so we sought a formal analytical proof of this conjecture:

$$(2 \phi_{STO} + 1 \phi_{STT}) /3 = \pi/(3\sqrt{2}) \quad . \tag{24}$$

Namely,

$$\frac{1}{3}\left(2 \frac{3}{4\sqrt{2}} \arccos\left(\frac{17}{81}\right) + 1 \sqrt{2} \arccos\left(\frac{23}{27}\right)\right) = \frac{\pi}{\sqrt{18}} \quad . \tag{25}$$

### 4.1. Stated as a theorem

"Let ϕ$_{STT}$ and ϕ$_{STO}$ denote the packing densities associated with the spherically truncated tetrahedron (STT) and the spherically truncated octahedron (STO), respectively, as defined by equations (18) and (28), then combining 1 ϕ$_{STT}$ with 2 ϕ$_{STO}$ has the packing density of CCP."

### 4.2. Proof

Multiplying both sides of equation (24) by $12\sqrt{2}$, we obtain the equivalent identity:

$$6 \arccos\left(\frac{17}{81}\right) + 8 \arccos\left(\frac{23}{27}\right) = 4\pi \quad . \tag{26}$$

Provide a geometric interpretation of this identity:

Consider a unit sphere (*i.e.*, a sphere of radius R = 1) having surface area A = 4π. We partition its surface into non-overlapping spherical polygons, as follows:

The term $6 \arccos\left(\frac{17}{81}\right)$ corresponds to the total area of six congruent *spherical squares* (see equation 16), each subtending a central angle of $\arccos\left(\frac{17}{81}\right)$.

The term $8 \arccos\left(\frac{23}{27}\right)$ corresponds to the total area of eight congruent *spherical triangles* (see equation 4), each subtending a central angle of $\arccos\left(\frac{23}{27}\right)$.

This partition arises naturally from the geometry of the cuboctahedron, a uniform Archimedean solid with 6 square and 8 triangular faces of equal edge length. When inscribed in a unit sphere, each face of the cuboctahedron defines a corresponding spherical polygon on the surface of the sphere.

Figure 10 illustrates this construction: the 6 spherical squares and 8 spherical triangles tessellate the entire surface of the sphere without overlap or gap. Since the sum of their areas equals the surface area of the sphere, it follows that:

$$6 \arccos\left(\frac{17}{81}\right) + 8 \arccos\left(\frac{23}{27}\right) = 4\pi$$

This verifies equation (26), from which equation (24) follows. This completes the proof.

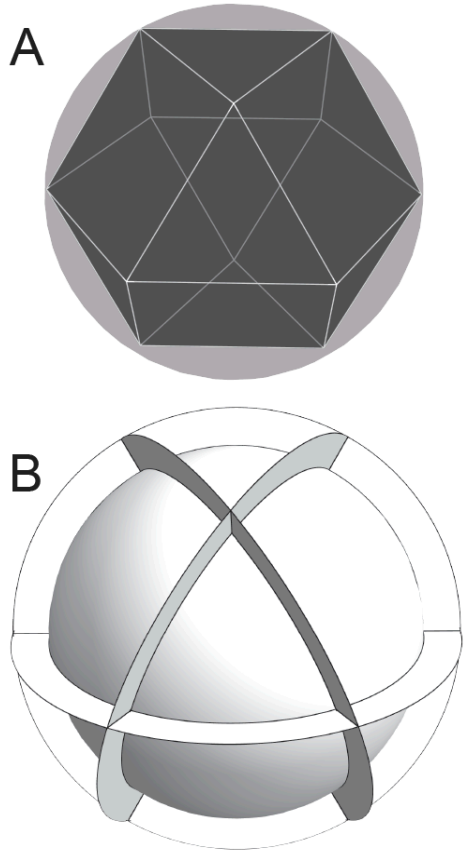

**Figure 10**. Cuboctahedron with faces of edge-length 1 unit in its circum-sphere of radius R = 1. **A**, Drawn using *Mathematica's* PolyhedronData["Cuboctahedron","Polygons"]. **B**, projection of the cuboactahedron onto its circum-sphere drawn in the interactive site https://demonstrations.wolfram.com/SphericalModelsOfPolyhedra/.

### 4.3. Theorem restated

Having now established the conjecture by analytical means, we formally state it as a theorem:

'The weighted average packing density of the spherical sections that define one sphero-planar octahedron (STT) and two sphero-planar cuboctahedra (STOs) from a cubic close packed array of uniform spheres, is equal to the packing density of the close packed spheres that generate these bodies in the first place, namely $\pi/\sqrt{18}$.'

## 5. Conclusions

Our study demonstrates that the combined volume occupied by eight STTs and six STOs comprises nearly 25% of the total volume of an unbounded CCP arrangement of uniform spheres. This result may seem counterintuitive, given that CCP achieves the highest possible packing density for identical spheres [7]. However, it follows from the significantly larger volume of the STO - approximately five times that of the STT - that such an imbalance between the constituent unit voids can still yield a substantial overall contribution to the total volume.

Several enticing directions arise from this work:

*1. Transport modelling* through STT–STO arrays using directed random walks, to study permeation properties in porous media;

*2. Extension of geometric modelling* via new sphero-planar polyhedra, generated using RegionPlot3D in *Mathematica*;

*3. Discovery of further identities*, akin to equations (24) and (25), that relate geometrical constructions to packing densities.

It is important to distinguish the theorem presented here from earlier results in three-dimensional tiling theory. Whereas classical tiling of Euclidean 3-space with polyhedra uses one regular octahedron and two regular tetrahedra of equal edge length [21], our result relates to a different geometrical configuration: namely, a weighted combination of two octahedral and one tetrahedral unit derived from CCP voids, yielding a mean packing density exactly equal to that of the original sphere array. This is a new and independent finding.

The initial goal of this investigation, namely, the definition and visualisation of two complementary void bodies (STT and STO) derived from a CCP lattice, was successfully achieved. Using GraphicsComplex in *Mathematica*, and placing the STT and STO units in their correct orientations and relative frequencies (4:3), we were able to generate a periodic structure that reproduces the original CCP array of spherical voids. This construction serves as a compelling demonstration of *Mathematica*'s capability in advanced geometric visualisation and symbolic computation and reflects broader recent advances in computational geometry.

The findings also challenge the oversimplified notion that the void space in disordered or random sphere packings can be adequately modelled by placing smaller spheres in the interstices [*e.g.*, 3]. The actual geometry of voids, particularly their surface-to-volume ratio, is far more nuanced. This has direct consequences in fields such as nuclear magnetic resonance (NMR), where surface-enhanced spin relaxation is sensitive to the local geometry of pores [*e.g.*, 12]. The internal surface area of the STT and STO units, as defined here, provides a more physically realistic basis for such modelling.

In summary, this work presents a systematic and geometrically rigorous approach to describing voids in sphere packings. It introduces a novel packing density identity for spheroplanar polyhedra derived from CCP and thereby offers new insight into the physical characterisation of porous materials. Our contribution complements and extends earlier qualitative accounts [*e.g.*, 2] by providing a quantitative framework rooted in explicit geometric and computational analysis.

## 6. Acknowledgements

I am grateful to Harry Calkins (Wolfram Research/Mathematica) for guidance on the use of GraphicsComplex, and to Louis Breton (also of Wolfram Research) for helpful advice regarding the projection of polyhedra onto their circumscribing spheres. Dr Daniel Daners, Applied Mathematics, University of Sydney, is thanked for valuable discussions. Support from the Australian Research Council under Discovery Project Grant DP190100510, concerned with modelling cell shapes in *Mathematica,* is gratefully acknowledged.

## 7. Mathematica Notebook

The *Mathematica* scripts used to generate the figures, and numerical results will be made available via the Wolfram Community website. The URL will be listed in the final version of the manuscript.